\documentclass{nature_fig}
\usepackage{graphicx} 
\usepackage{xcolor}
\usepackage{amsmath}
\usepackage{bm}
\usepackage[sort&compress,super,comma,sectionbib]{natbib}
\usepackage{chapterbib}
\usepackage[font={small},labelfont={bf}]{caption}
\usepackage{txfonts}
\usepackage{pdfpages}

\usepackage{lineno}
\modulolinenumbers[1]

\newcommand{\Cs}{{CsV$_3$Sb$_5$}}
\newcommand{\K}{{KV$_3$Sb$_5$}}

\newcommand{\all}{{\textit{A}V$_3$Sb$_5$}}
\newcommand{\Tcdw}{{$T_{\rm CDW}$}}

\newcommand{\Kerr}{{$\theta_{\rm K}$}}
\newcommand{\DKerr}{{$\Delta\theta_{\rm K}$}}

\title{Time-reversal symmetry breaking in charge density wave of \Cs\ detected by polar Kerr effect
}

\author{
Yajian~Hu$^{1\ast}$, Soichiro~Yamane$^{1}$, Giordano~Mattoni$^{1,2}$, Kanae Yada$^{1}$, Keito Obata$^{1}$, Yongkai~Li$^{3,4,5}$, Yugui~Yao$^{3,4}$, Zhiwei~Wang$^{3,4,5}$, Jingyuan~Wang$^{6}$, Camron~Farhang$^{6}$, Jing~Xia$^{6}$, Yoshiteru~Maeno$^{1,2}$, Shingo~Yonezawa$^{1\dagger}$
}

\begin{document}

\maketitle

\vspace{-0.5cm} 

\begin{affiliations}
{\small
 \item Department of Physics, Graduate School of Science, Kyoto University, Kyoto 606-8502, Japan
 \item Toyota Riken-Kyoto University Research Center (TRiKUC), Kyoto University, Kyoto 606-8501, Japan
 \item Centre for Quantum Physics, Key Laboratory of Advanced Optoelectronic Quantum Architecture and Measurement (MOE), School of Physics, Beijing Institute of Technology, Beijing 100081, P. R. China
 \item Beijing Key Lab of Nanophotonics and Ultrafine Optoelectronic Systems, Beijing Institute of Technology, Beijing 100081, P. R. China
 \item Material Science Center, Yangtze Delta Region Academy of Beijing Institute of Technology, Jiaxing 314011, P. R. China
 \item Department of Physics and Astronomy, University of California, Irvine, California 92697, USA

}
\end{affiliations}

\vspace{-0.8em}

\noindent
$^{\ast}$e-mail: hu.yajian.78e@st.kyoto-u.ac.jp
$^{\dagger}$e-mail: yonezawa@scphys.kyoto-u.ac.jp

\vspace{-1.5em}

\noindent \textit{Dated: \today}

\vspace{1em}


\begin{abstract}
The Kagome lattice exhibits rich quantum phenomena owing to its unique geometric properties.
Appealing realizations are the Kagome metals \all\ (\textit{A} = K, Rb, Cs), where unconventional charge density wave (CDW) is intertwined with superconductivity and non-trivial band topology. Several experiments suggest that this CDW is a rare occurrence of chiral CDW characterized by orbital loop current. 
However, key evidences of loop current, spontaneous time-reversal symmetry-breaking (TRSB) and the coupling of its order parameter with the magnetic field remain elusive. 
Here, we investigate the CDW in \Cs\ by magneto-optic polar Kerr effect with sub-microradian resolution. Under magnetic field, we observed a jump of the Kerr angle at the CDW transition. This jump is magnetic-field switchable and scales with field, indicating magneto--chirality coupling related to non-trivial band topology. At zero field, we found non-zero and field-trainable Kerr angle below \bm{$T_{\rm CDW}$}, signaling spontaneous TRSB. Our results provide a crucial step to unveil quantum phenomena in correlated Kagome materials.

\end{abstract}


The Kagome lattice has gained the spotlight over decades because of its rich ground states owing to its intriguing band structure. The nearest-neighbor tight-binding model of a two-dimensional (2D) Kagome net depicts presence of a flat band, Dirac cones, and van Hove singularities (vHSs).
The flat band with localized electrons gives rise to strongly correlated phenomena, while the Dirac cones assist systems to become topological insulators or Weyl semimetals \cite{kang2020dirac, liu2018Co3Sn2S2}. The vHSs, showing divergent density of states, can lead to electronic instability such as spin-density wave and charge-density wave (CDW).

The Kagome metals \all\ (\textit{A} = K, Rb, Cs) have attracted tremendous interest recently \cite{Ortiz2019}.
Crystallographically, vanadium atoms form an ideal 2D Kagome net at ambient temperature (Fig.~\ref{fig1}a, b). In the band structure, traces of the flat band, Dirac cones, and vHSs expected for a 2D Kagome net all appear \cite{Ortiz2020Cs}. Both the Dirac cone and vHSs are located close to the Fermi level, promoting the topological and strongly correlated phenomena. 
\all\ compounds display CDW order and superconductivity \cite{Ortiz2019, Ortiz2020Cs, Ortiz2021K, Yin2021Rb}. Moreover, giant anomalous Hall effect (AHE) is reported to occur concurrently with the CDW transition in \K\ \cite{Yang2020aheK} and \Cs\ \cite{Yu2021aheCs}.
Angle-resolved photoemission spectroscopy (ARPES) \cite{kang2022twofold, Nakayama2021,wang2021distinctive,luo2022cdwgap} and X-ray scattering \cite{Li2021Xray} experiments show a strongly momentum-dependent CDW gap and indicate that the CDW transition is triggered by Fermi surface nesting dominated by vHSs. 
In the context of lattice distortions in the CDW state, the acoustic phonon anomaly is absent \cite{Li2021Xray} while the hardening of an optical phonon mode is reported \cite{Xie2022INS}.
These results establish the unconventional nature of the CDW in the normal-state and it can profoundly influence the superconducting state, which also exhibits intriguing features \cite{Xu2021sc, zhao2021nodal, Chen2021roton, mielke2022time, Liang2021cdw3q, Ni2021anisotropic}. Thus, \all\ provide a versatile platform to investigate the interplay between novel band topology, CDW and superconductivity in the Kagome lattice.

A critical issue of the CDW in \all\ concerns the accompanied symmetry breakings. In scanning tunnelling microscopy (STM) studies, three-dimensional $2\times2\times2$ or $2\times2\times4$ charge modulations have been mapped out \cite{Liang2021cdw3q, Ortiz2021QO}. Interestingly, clockwise or counterclockwise intensity modulation of putatively identical superlattice peaks exist and can be flipped by magnetic field, as reported for all members of the \all\ family \cite{Jiang2021stmK, Zhiwei2021stmCs, Shumiya2021Rb}. These results raise the possibility of time-reversal symmetry breaking (TRSB) although the long-range magnetic order is absent \cite{Ortiz2019}.
Zero-field muon spin relaxation ($\mu$SR) on \K\ and \Cs\ show enhanced relaxation rates in the CDW states, supporting the TRSB \cite{mielke2022time, yu2021musr, Khasanov2022}. 
On the other hand, an additional $C_6$ rotation symmetry breaking is found at a lower temperature in the CDW state and the electronic ground state is proposed to be nematic \cite{Li2022rotation, Jiang2021stmK, Ratcliff2021, Wang2021pumpprobe, Shumiya2021Rb, Xiang2021nematic, zhao2021cascade}. 
Theoretically, the TRSB and rotation symmetry breaking can be well-explained by orbital loop currents and chiral flux induced as a consequence of order parameters with non-zero imaginary component.
Such an unconventional CDW state is now called chiral CDW. \cite{Denner2021analysis, Feng2021theory, Feng2021chiral, Lin2021theory}. However, several experiments report the absence of TRSB in the CDW phase \cite{kenney2021absence,song2022orbital,Li2022nochiral,Li2022rotation}, and theories propose the charge bond order with real order parameters \cite{Tazai2022, Park2021}. We comment that the charge distribution in the CDW state can have a structural chirality \cite{Park2021}, which should not be confused with the chirality and TRSB due to loop current. For a thorough understanding of the CDW in \all, new probes of TRSB are crucial.

Here, we performed high resolution magneto-optic Kerr effect (MOKE) measurements to investigate the CDW in \Cs, a canonical member of the \all\ family. We measured the polar Kerr effect (PKE) using a zero-loop Sagnac interferometer. Different from the ordinary optical setup, PKE measured with our setup is only sensitive to TRSB of the system \cite{Xia_APL}.
This technique has been used for studying novel TRSB phases in quantum materials, for instance, the chiral SC of Sr$_2$RuO$_4$ \cite{Xia2006sro} and uranium-based superconductors \cite{Schemm2014UPt3}, and the chiral charge order in the high-$T_{\rm{c}}$ cuprates \cite{Karapetyan2014pke}.
Using a newly-built all-fiber Sagnac interferometer, we measured polar Kerr angle \Kerr\ at low temperature with sub-microradian resolution, with and without magnetic field. 
Under magnetic field, the polar Kerr angle jump \DKerr\ at the CDW transition is almost linear in field and changes sign with opposite field direction. We demonstrate that this strong coupling to magnetic field is related to non-trivial band topology.
At zero field, non-zero \Kerr\ has been detected below \Tcdw. This spontaneous \Kerr\ is switchable by out-of-plane magnetic fields applied while cooling across the CDW transition, implying spontaneous TRSB in \Cs. Our results can be well-interpreted by a magnetic-field switchable loop-current order in the chiral CDW.

\section*{Results}
\begin{figure}
\begin{center}
\includegraphics[width=16cm]{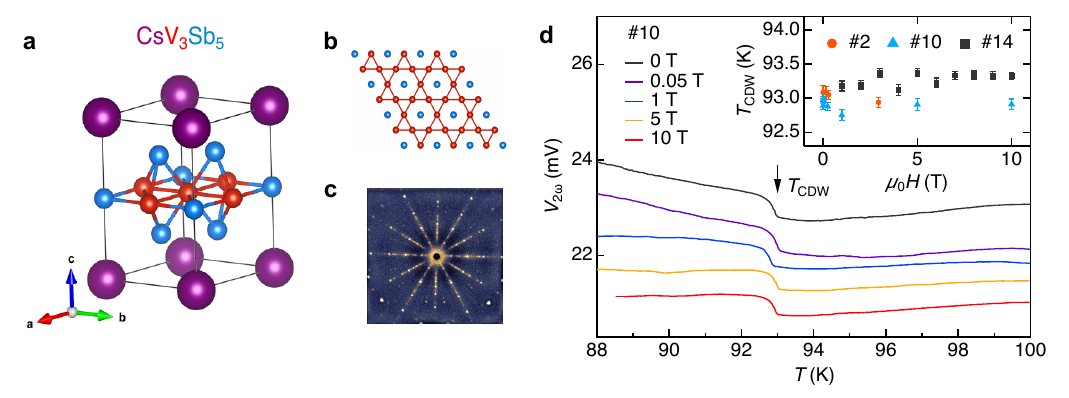}
\end{center}
\caption{
{\bf Crystal structure of \Cs\ and CDW transition detected by reflectivity.}
{\bf a}. Room-temperature crystal structure of \Cs. 
{\bf b}. Kagome net formed by V atoms (red spheres). Sb atoms (blue spheres) are located at the hexagonal center. The crystal structure is illustrated by using the software VESTA \cite{Vesta}.
{\bf c}. Laue diffraction image taken on sample $\#$14 along the [001] direction. 
{\bf d}. Temperature dependence of the second-harmonic signal $V_{2\omega}$, which is proportional to the sample reflectivity, near the CDW transition measured on sample $\#$10 at different magnetic fields. The arrow indicates the CDW transition. The inset shows the magnetic-field dependence of the transition temperature \Tcdw\ of samples $\#$2, $\#$10 and $\#$14. \Tcdw\ is determined by the peak in the temperature derivative of $V_{2\omega}$. These data show that \Tcdw\ is nearly the same for all three samples and independent of \textit{H}. 
\label{fig1}
}
\end{figure}

\noindent{\bf Crystal characterization and CDW transition in reflectivity.}
Before optical measurements, the samples were characterized by Laue diffraction and magnetic susceptibility. The Laue diffraction image shown in Fig.~\ref{fig1}c displays clear spots, consistent with the hexagonal ($P6/mmm$) crystal structure at room temperature and indicating high crystal quality without significant stacking faults. The magnetic susceptibility shows sharp transitions at $\sim94$~K and $\sim2.9$~K, corresponding to the CDW and the superconducting transitions, respectively (see Supplementary Fig.~S1). These data are consistent with previous reports \cite{Ortiz2019, Ortiz2020Cs} and demonstrate that the samples are well single-crystalline.

In our PKE experiments, we used a broadband light source with center wavelength of $\sim1550$~nm, and the phase of the light is modulated by the phase modulator with the frequency $\omega/2\pi\sim3.848$~MHz. We detect the first and second harmonic components of the reflected signal from a photo detector \cite{Xia2006sro}. The second harmonic signal $V_{2\omega}$ is proportional to the reflectivity of the sample, while the ratio between the first harmonic signal $V_{1\omega}$ to $V_{2\omega}$ gives the polar Kerr angle \Kerr\ $\approx0.271V_{1\omega}/V_{2\omega}$ (See Methods).
Figure~\ref{fig1}d displays the temperature dependence of $V_{2\omega}$ for several selected magnetic fields between 0~T and 10~T. Upon cooling, the reflectivity jumps and increases below $\sim93$~K. Because of the gap opening at the CDW transition, the spectral weight of low-energy optical conductivity shifts to higher energy ranges. This leads to the increase in the reflectivity at our wavelength, $\sim1550$~nm (0.8~eV), as observed by the optical spectroscopy \cite{Zhou2021optic,Uykur2021Cs}. From the temperature derivative of $V_{2\omega}$ \cite{SUPP}, we extract the transition temperature \Tcdw\ at different magnetic fields and plot it in the inset of Fig.~\ref{fig1}d. Among the three samples that we have measured, \Tcdw\ is located within the range $93\pm0.5$~K, and remains independent of the external field up to 10~T.

\noindent{\bf Polar Kerr angle measurements under magnetic field.}
We first present the polar Kerr angle \Kerr\ under magnetic field. Because of the Faraday effect of the lens, there is a background signal contributing to the measured \Kerr. In order to examine the signal from the sample across \Tcdw, we use the measured \Kerr\ of a Nb metal sheet to subtract the background signal $\theta_{\rm K}^{\rm bg}$ (see Supplementary Fig.~S2). Figure~\ref{fig2}a shows extracted sample signal $\theta_{\rm K}^{\rm sample}=\theta_{\rm K}-\theta_{\rm K}^{\rm bg}$ divided by magnetic field. At \Tcdw, $\theta_{\rm K}^{\rm sample}$ jumps and gradually increases with decreasing temperature, consistent with a first-order transition as previously reported \cite{Ortiz2020Cs}. Moreover, with opposite magnetic field, $\theta_{\rm K}^{\rm sample}$ changes its sign to negative. 
The fact that $\theta_{\rm K}^{\rm sample}$ rises concurrently with the CDW transition, and reverses its sign with opposite magnetic field direction is in agreement with the intensity-modulation reversal observed in STM spectroscopy \cite{Zhiwei2021stmCs} and the AHE \cite{Yu2021aheCs}, which are interpreted in terms of TRSB and large Berry phase in the CDW state of \Cs.

Figure~\ref{fig2}b shows the temperature dependence of the measured polar Kerr angle \Kerr. With increasing magnetic field, the magnitude of \Kerr\ and the jump \DKerr\ at \Tcdw\ both increase. 
In order to obtain the field dependence of the polar Kerr angle, we determine \DKerr\ from the difference between two linear fits close to \Tcdw\ (the inset of Fig.~\ref{fig2}c) and plot it as a function of magnetic field (Fig.~\ref{fig2}c). For all three samples, \DKerr\ is almost linear in magnetic field in the studied range between -6~T and +10~T, indicating the polar Kerr angle in the CDW state is strongly coupled to the magnetic field. Hence, it is natural to compare \DKerr, proportional to the Hall conductivity at the optical frequency, with magnetization. For the conventional anomalous Hall conductivity, one expects its magnitude to be proportional to the magnetization. 
\Cs\ is paramagnetic and the magnetization decreases below \Tcdw\ due to the decreasing density of states (DOS) (see Supplementary Fig.~S1). If \DKerr\ originates from the change of magnetization, we would expect also \Kerr\ to decrease at \Tcdw. Our data is opposite to this expectation. 

\noindent{\bf Correct sign of the Kerr angle.}
We have now realized that the polar-Kerr-effect results under magnetic field shown in the first submission of this manuscript display an opposite sign due to incorrect calibration of the previous setup. In that setup, the signal strength depended non-linearly on the input power. With the new setup, the signal varies linearly on power, and the sign is calibrated with ferromagnetic SrRuO$_3$. The polar Kerr angle jump at \Tcdw\ has a negative sign, as shown in Fig.~\ref{figc1} below. We note that the magnitude of the polar Kerr angle jump is almost the same as that in our first report; the magnitude as well as the sign are consistent with Saykin \textit{et al.} (arXiv:2209.10570) and Wang \textit{et al.} (arXiv:2301.08853).
Concerning the Kerr angle after training under various fields, we confirmed signals suggesting the time reversal symmetry breaking (TRSB) in the crystal ($\#$14) exhibiting a very sharp CDW transition, as shown in Fig.~\ref{figc1}. For other sample crystals with a broader CDW transition, we did not observe a signal suggestive of TRSB.
Note that in this version, Fig. 3, Fig. 4a-c and Fig S6, as well as the relevant text in v1 have been removed.

\noindent{\bf Spontaneous TRSB in the CDW phase evidenced by zero-field Kerr effect.}
At last, we discuss the presence of non-zero polar Kerr angle at zero field, key feature of spontaneous TRSB and chiral CDW in \Cs. 
The data collected at zero field shows \Kerr\ fluctuates around zero by $\pm0.2$~µrad at high temperature and becomes finite below \Tcdw, indicating the spontaneous TRSB in the CDW state. 
We observed that the magnitude and the sign of \Kerr\ vary for each measurement. We repeated this measurement with different optical power and observed similar behaviour.
This can be well-explained by random formation of TRSB domains after zero-field-cooling (ZFC). In addition, the magnitude of \Kerr\ tends to grow at a temperature lower than \Tcdw, probably due to the growth of domain sizes towards low temperature. 

In order to obtain clearer TRSB signal and examine the presence of TRSB domains, we performed “field training” experiments to align them. We cooled the sample across \Tcdw\ under a perpendicular magnetic field, set the field to zero at low temperature, and then collected data upon zero-field-warming (ZFW). 
At $T>T_{\rm CDW}$,  the data fluctuate around zero. 
The successful field training implies that the domains are aligned by the magnetic field and that a residual moment persists after removing the field.
Moreover, the magnitude of \Kerr\ without training is in the range $\pm1$~µrad, while \Kerr\ after field training reaches $\sim2$~µrad in some runs at low temperature. This shows how field training effectively aligns TRSB domains. 
The non-zero \Kerr\ and the training effect can be well interpreted as a consequence of TRSB in the CDW state, possibly with a transition temperature $T_{\rm TRSB}\approx$~\Tcdw.
%

\section*{Discussion}
The non-zero \Kerr\ at zero field arising below \Tcdw\ is a sign of spontaneous TRSB in the CDW state.
Since long-range magnetic order has not been detected in \Cs, the TRSB is not due to magnetism. 
Instead, we discuss below that the observed TRSB can be explained in terms of loop-current order in the CDW state.

The CDW in \Cs\ is dominated by Fermi-surface nesting characterized by wavevectors connecting the vHSs at the M points of the hexagonal Brillouin zone. The electronic states around the vHS are dominated by vanadium $d$ orbitals. 
Due to the three sublattices of the Kagome net, the inhomogeneous DOS at $E_{\rm F}$ promotes nearest-neighbor interaction, thereby resulting in periodic modulation of the bond strength instead of the onsite charge density. This corresponds to a charge bond order (CBO). Further Ginzburg--Landau analysis shows that the three order parameters are imaginary with unequal phases \cite{Denner2021analysis, Feng2021theory, Feng2021chiral, Lin2021theory}. The imaginary CBO spontaneously induces orbital loop current in the Kagome lattice and hence breaks time-reversal symmetry, as sketched in Fig.~\ref{fig4}. The loop currents (gray arrows) generate opposite chiral flux in hexagons (blue arrow) and triangles (orange arrows). Since both the bond and on-site charge densities are inhomogeneous, an uncompensated net flux appears in each unit cell. 
Different chiral CDW patterns form domains, and thus, the residue moments over domains are responsible for non-zero polar Kerr angle at zero field. These domains can be aligned by a moderate magnetic field and result in the observed field-training effect.

In \all, competing phases have been proposed. The CBO theory with imaginary order parameters suggests rotational-symmetry breaking without translational-symmetry breaking, namely electronic nematicity in the CDW phase \cite{Denner2021analysis}. Such nematicity may be the reason of the bulk two-fold rotational symmetry observed in transport measurements \cite{Li2022rotation, Xiang2021nematic}. 
The CBO theories for the CDW are based on the two-dimensional Kagome lattice. Remarkably, the interlayer charge modulation appears to be important in \Cs. STM measurement showed that a $\pi$-phase shift of the charge modulation intensity across single-unit-cell surface steps has three-dimensional ordering \cite{Liang2021cdw3q}. Out-of-plane superlattice peaks have also been observed in X-ray diffraction \cite{Chen2022xray, Stahl2021, Li2021spatial, Xiao2022xray}. 
Such interlayer coupling may lead to the emergence of multiple sub-phases in the CDW.
Indeed, high pressure experiments showed suppression of the CDW phase and unusual two-peak superconducting dome at a moderate pressure \cite{Yu2021pressure, Chen2021pressure}. 
In a µSR study of \Cs\ \cite{yu2021musr}, TRSB was observed below 70~K (noticeably lower than \Tcdw$\sim 94$~K) along with an additional magnetic transition below 30~K. Other µSR studies on \Cs\ and \K\ report that TRSB occurs just below \Tcdw\ \cite{mielke2022time, Khasanov2022}. 
In our data of \Kerr\ measured at zero field, we could not resolve clear transitions between multiple phases, but some datasets exhibit features such as slope or intensity changes at around 60~K and 80~K.  
These features may hint at competing phases in CDW and call for further investigation.

We notice that optical polarization rotation studies on \Cs\ were performed very recently in zero field using 800~nm light and by ordinary polarizer-based optical setups, methodologies different from ours \cite{Wu2021Kerr, Xu2022_Kerr}. Both works show the emergence of birefringence and isotropic polarization rotation below \Tcdw. The former indicates the electronic nematicity and the latter may come from either TRSB or structural chirality. 
In our study, we use the zero-loop Sagnac interferometer and the detected polar Kerr angle is solely due to TRSB \cite{Xia_APL}. Furthermore, our optical results under magnetic field and field-training effect show that the polar Kerr effect in \Cs\ is strongly coupled to magnetic field.

In conclusion, we have studied the TRSB in CDW state of \Cs\ via polar Kerr effect.
Under magnetic field, the Kerr-angle jump at the CDW transition \DKerr\ is proportional to the magnetic field and its sign is field-switchable. 
At zero field, non-zero \Kerr\ arises in the CDW state and can be effectively trained by magnetic field. These results point toward spontaneous TRSB in the CDW states, which can be explained by the loop-current order theory. 
Our polar Kerr effect results elucidate the central issue of the symmetry breaking in the CDW of \Cs, motivating further experimental and theoretical research on these Kagome metals. 
Our study demonstrates that the high-resolution MOKE experiment under low temperature and high magnetic field condition is a powerful tool for exploring non-trivial symmetry breakings, which is crucially important for exploring exotic quantum materials.

\begin{methods}
High-quality single crystals of \Cs\ were grown by a self-flux method with Cs-Sb binary eutectic mixture (Cs$_{0.4}$Sb$_{0.6}$) as flux \cite{Zhiwei2021stmCs}. The typical lateral size of the obtained crystal is $\sim3$~mm. The crystal structure were examined by Laue diffraction at room temperature. The magnetization was measured by a SQUID magnetometer (MPMS, Quantum Design). Polar Kerr angle was measured using an all-fiber zero-loop Sagnac interferometer, constructed based on Ref.~\citenum{Xia2006sro}. We used a broadband light source with wavelength centered at 1550~nm and a line width of 110~nm. The incident light was split into horizontally and vertically polarized components of the polarization-maintaining (PM) fiber, and then phase modulated using an electro-optical modulator (EOM). The modulation frequency was $\omega/2\pi$ $\sim3.848$~MHz so that the light reflected from the sample has a $\pi$ phase shift relative to the incident light. The maximum incident optical power is 200~$\mu$W and the spot size is $\sim4~\mu$m. 
The sample surface was cleaved before optical measurements, with typical thickness after cleavage of $\sim100~\mu$m. The low-temperature measurements were performed in a $^4$He refrigerator with a base temperature of about 1.9~K . The magnetic field was applied by a 11~T superconducting magnet along the $\textit{c}$ axis. In the field training measurements, a Hall sensor was used to determine the zero-field with an error within 5~Oe.

\end{methods}

\bibliography{CsV3Sb5_CDW}

\section*{Data Availability}
All the data that support the findings of this paper are available from the corresponding authors upon reasonable request.

\section*{Acknowledgments}
We acknowledge M. Cuoco, A. Kapitulnik, Y. Yanase, and J. Goryo, for fruitful discussion. 
This work was supported by Grant-in-Aids for
Scientific Research on Innovative Areas ``Quantum Liquid Crystals'' (KAKENHI Grant Nos. 20H05158, 22H04473) from the Japan Society for the Promotion of Science (JSPS),
a Grant-in-Aid for JSPS Fellows (KAKENHI Grant No. 20F20020) from JSPS,
Grant-in-Aids for Scientific Research (KAKENHI Grant Nos. 17H06136, 22H01168) from JSPS,
by Core-to-Core Program (No. JPJSCCA20170002) from JSPS, 
by a research support funding from The Kyoto University Foundation,
and by ISHIZUE 2020 of Kyoto University Research Development Program.
G. Mattoni acknowledges the support from the Dutch Research Council (NWO) through Rubicon Grant No. 019.183EN.031.
The work at Beijing was supported by the Natural Science Foundation of China (Grant No. 92065109), the National Key R$\&$D Program of China (Grant No. 2020YFA0308800), the Beijing Natural Science Foundation (Grant No. Z210006). Z. Wang thanks the Analysis$\&$Testing Center at BIT for assistance in facility support. 
The work at UCI is funded by the Gordon and Betty Moore Foundation through Grant GBMF10276 to Jing Xia.

\section*{Author contributions}
This study was designed by S. Yonezawa and Y. Maeno. 
Y. Hu and S. Yamane performed polar Kerr effect measurements and analyses, with assistance from G. Mattoni, K. Yada, J. Wang, C. Farhang, J. Xia, S. Yonezawa, and Y. Maeno. Y. Hu, S. Yamane, and K. Obata performed the magnetization measurements. K. Obata took Laue diffraction images.
Z. Wang, Y. Li, and Y. Yao grew single crystalline samples and characterized them. The manuscript was mainly written by Y. Hu and S. Yonezawa, based on discussion among all authors.

\section*{Competing financial interests}

All authors declare there is no competing interests regarding this work.
\clearpage


\clearpage

\begin{figure}
\begin{center}
\includegraphics[width=16cm]{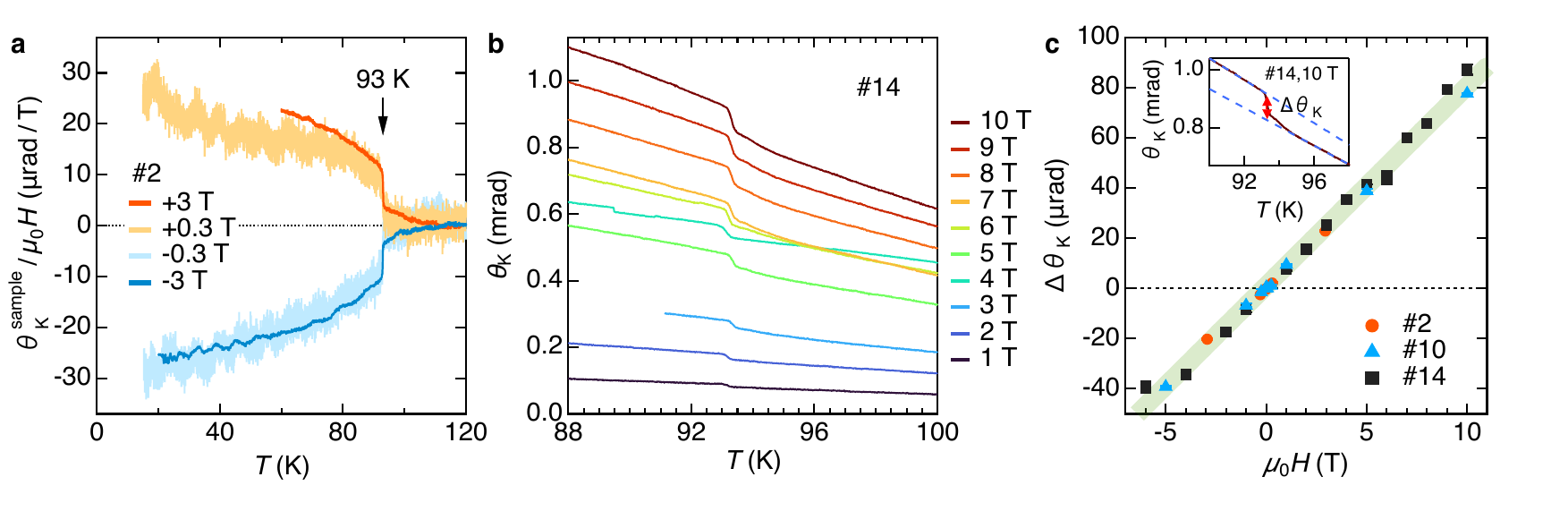}
\end{center}
\caption{{\bf The sign of polar Kerr angle in these figures is incorrect; please refer to Fig.3 below.}
{\bf Temperature and magnetic field dependence of the polar Kerr angle of \Cs.}
{\bf a}. Temperature dependence of the background-subtracted polar Kerr angle $\theta_{\rm K}^{\rm sample}=\theta_{\rm K}-\theta_{\rm K}^{\rm bg}$ divided by magnetic field for sample $\#$2. The polar Kerr angle \Kerr\ of a Nb metal sheet obtained by a separated measurement was used as the background $\theta_{\rm K}^{\rm bg}$ (see Supplementary Fig.~S2). 
{\bf b}. Temperature dependence of \Kerr\ for $\#$14 measured from 1~T (bottom curve) to 10~T (top curve) upon cooling. 
{\bf c}. Polar Kerr angle jump \DKerr\ at \Tcdw\ as a function of magnetic field for three samples. \DKerr\ is determined by the difference at \Tcdw\ between the linear fits below and above \Tcdw, as examplified in the inset. The shaded region indicates a line with slope $\Delta\theta_{\rm K}/\mu_0H\sim8.0$~$\mu$rad/T deduced from all the data.
\label{fig2}
}
\end{figure}

\clearpage


\begin{figure}
\begin{center}
\includegraphics[width=10cm]{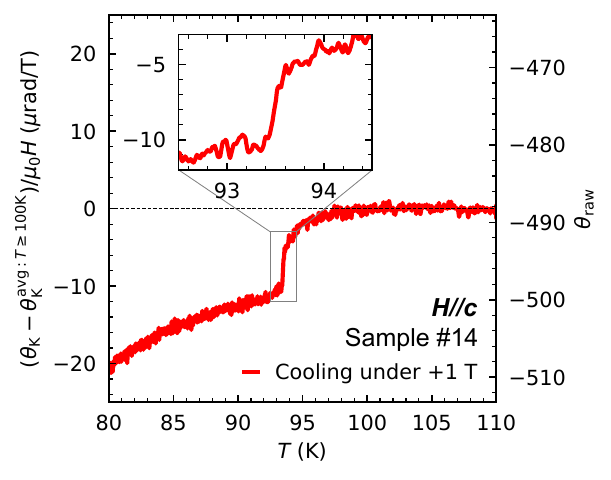}
\end{center}
\caption{
{\bf Change of the polar Kerr angle of a \Cs\ crystal near its CDW transition temperature under a c-axis magnetic field of 1~T.}
The averaged Kerr angle between 100 and 110~K has been subtracted to emphasize the sample contribution in the CDW state. As shown in the inset, this crystalline sample ($\#$14) exhibits a negative jump at \Tcdw\ of -5~$\mu$rad/T with a width as small as 0.2~K. The jump for this sample ($\#$14) is much sharper than other samples that we measured and also than those reported by other groups.
\label{figc1}
}
\end{figure}
\clearpage

\begin{figure}
\begin{center}
\includegraphics[width=10cm]{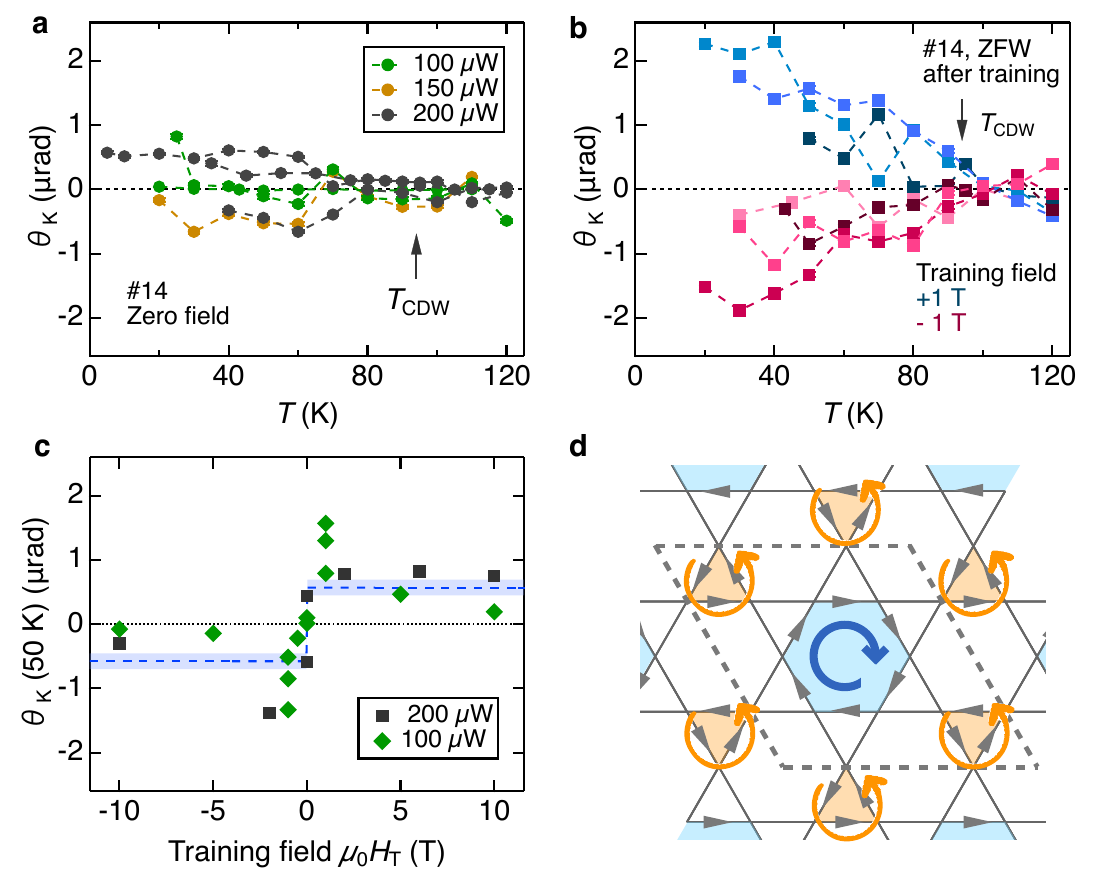}
\end{center}
\caption{
\linespread{1.5}\selectfont{}
Schematic of the real-space loop-current pattern in the Kagome lattice \cite{Lin2021theory}. The grey arrows indicate the orbital current direction. The associated chiral flux is indicated by the clockwise and counter-clockwise arrows. Within the unit cell in the CDW state (dashed lines), there are two triangles and one hexagon with opposite chiral flux. Their incomplete compensation, predicted by theories, would result in non-zero spontaneous Kerr effect.
\label{fig4}
}
\end{figure}
\clearpage



\section*{Supplementary material}

**Magnetic susceptibility of CsV$_3$Sb$_5$**

Figure S1a shows the temperature dependence of the magnetic susceptibility $\chi = M/H$ measured under in-plane field of 7 T for both zero-field-cooling (ZFC) and field-cooling (FC) conditions. It shows paramagnetic behaviour at high temperature. The magnetic susceptibility $\chi$ shows a sharp drop at $\sim 94$ K, corresponding to the charge density wave (CDW) transition. At $T_{\text{CDW}}$, gaps open on several parts of the Fermi surface and hence the density of states (DOS) at the Fermi level drops, leading to the decrease of the Pauli susceptibility. The ZFC and FC data almost overlap with each other. Figure S1b is the magnetic susceptibility measured at 5 Oe below 5 K. A sharp superconducting transition is observed at $T_c^{\text{onset}} \sim 2.9$ K. These data are consistent with previous studies[1,2] and indicative of high quality of the sample.

**Background subtraction for CsV$_3$Sb$_5$ sample #2**

In our experiment, the Faraday effect of lens and fiber becomes significant under magnetic field and contribute to the measured signal. In order to quantify the polar Kerr effect from the CDW transition in CsV$_3$Sb$_5$, we used a Nb metal sheet, which is non-magnetic above $T_c \sim 9.2$ K, to measure the background contribution. This measurement was done under experimental conditions similar to the measurements of sample #2. Figure S2a shows the measured $\theta_{\text{K}}$ for Nb at 0.3 T. We fit the data with a polynomial function to capture the temperature dependence and the resulted curve is denoted as $\theta_{\text{K}}^{\text{Nb}}$. The background signal at a different magnetic field $H$ is defined as $\theta_{\text{K}}^{\text{bg}} = \theta_{\text{K}}^{\text{Nb}} \times \frac{\mu_0 H}{0.3} \times C$, where $C$ is a constant determined by matching $\theta_{\text{K}}^{\text{bg}}$ with the data of CsV$_3$Sb$_5$ above

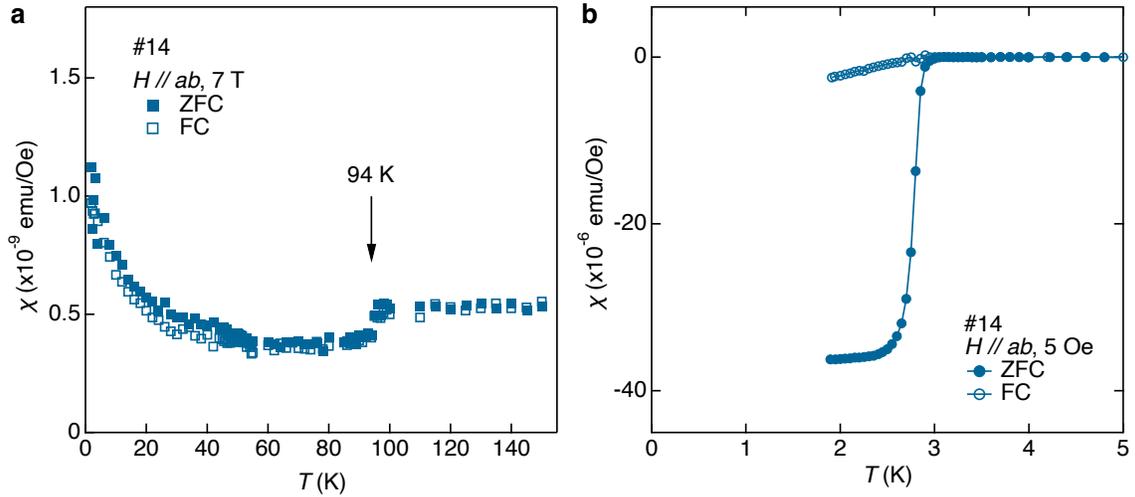

**Figure S1: Temperature dependence of the magnetic susceptibility of CsV$_3$Sb$_5$ of sample #14. a**. Temperature dependence of the magnetic susceptibility $\chi$ measured at 7 T. The magnetic field is in the *ab* plane. The closed markers represent data collected at warming after zero-field-cooling (ZFC), while the open markers represent data measured during field-cooling (FC). The arrow indicates $T_{CDW} \sim 94$ K. **b**. Temperature dependence of $\chi$ measured at 5 Oe parallel to the *ab* plane below 5 K, showing a sharp superconducting transition. The solid markers represent data collected at warming after zero-field-cooling (ZFC). The open markers represent data measured during field-cooling (FC).

$T_{\mathrm{CDW}}$. We found that the constant $C$ deviates from 1 typically by about 20%. This may be due to the slight changes in the modulation frequencies and the positions of the lens and fiber in the magnetic field. Figures S2b and c show $\theta_{\mathrm{K}}$ of $CsV_3Sb_5$ measured at 0.3 T and 3 T together with the background signal $\theta_{\mathrm{K}}^{\mathrm{bg}}$. By subtracting $\theta_{\mathrm{K}}^{\mathrm{bg}}$ from $\theta_{\mathrm{K}}$, we obtain the background-subtracted polar Kerr angle $\theta_{\mathrm{K}}^{\mathrm{sample}} = \theta_{\mathrm{K}} - \theta_{\mathrm{K}}^{\mathrm{bg}}$, as shown in Fig. S2d and Fig. 2a in the main text. Note that in Fig. S2c, the measured $\theta_{\mathrm{K}}$ and $\theta_{\mathrm{K}}^{\mathrm{bg}}$ start to deviate below $\sim 100$ K and make $\theta_{\mathrm{K}}^{\mathrm{sample}}$ at 3 T above $T_{\mathrm{CDW}}$ increase. This is probably because the $\theta_{\mathrm{K}}$ of Nb was measured at 0.3 T and may have slightly different temperature dependence from $\theta_{\mathrm{K}}$ at 3 T. Nevertheless, the small upturn in the $\theta_{\mathrm{K}}^{\mathrm{sample}}$ at 3 T does not affect our observation that the polar Kerr angle from the sample changes at $T_{\mathrm{CDW}}$ and this change can be switched by magnetic field.

**Polar Kerr angle, second-harmonic signal and their temperature derivatives**

In this section, we present the raw data used to extract the transition temperature $T_{\mathrm{CDW}}$ and polar Kerr angle jump $\Delta\theta_{\mathrm{K}}$, presented in the inset of Fig. 1b and Fig. 2c of the main text. Among the three measured samples, the data are consistent.

Figure S3a shows the temperature derivative of the second-harmonic signal $V_{2\omega}$ (proportional to the sample reflectivity) near the CDW transition of #14. The data are collected from 1 T to 10 T. The sharp peaks around 93 K indicate the CDW transition. $T_{\mathrm{CDW}}$ is determined by the temperature where $|dV_{2\omega}/dT|$ peaks.

Figure S3b shows $\theta_{\mathrm{K}}$ measured under negative magnetic field. At $T_{\mathrm{CDW}}$, the polar Kerr angle

4

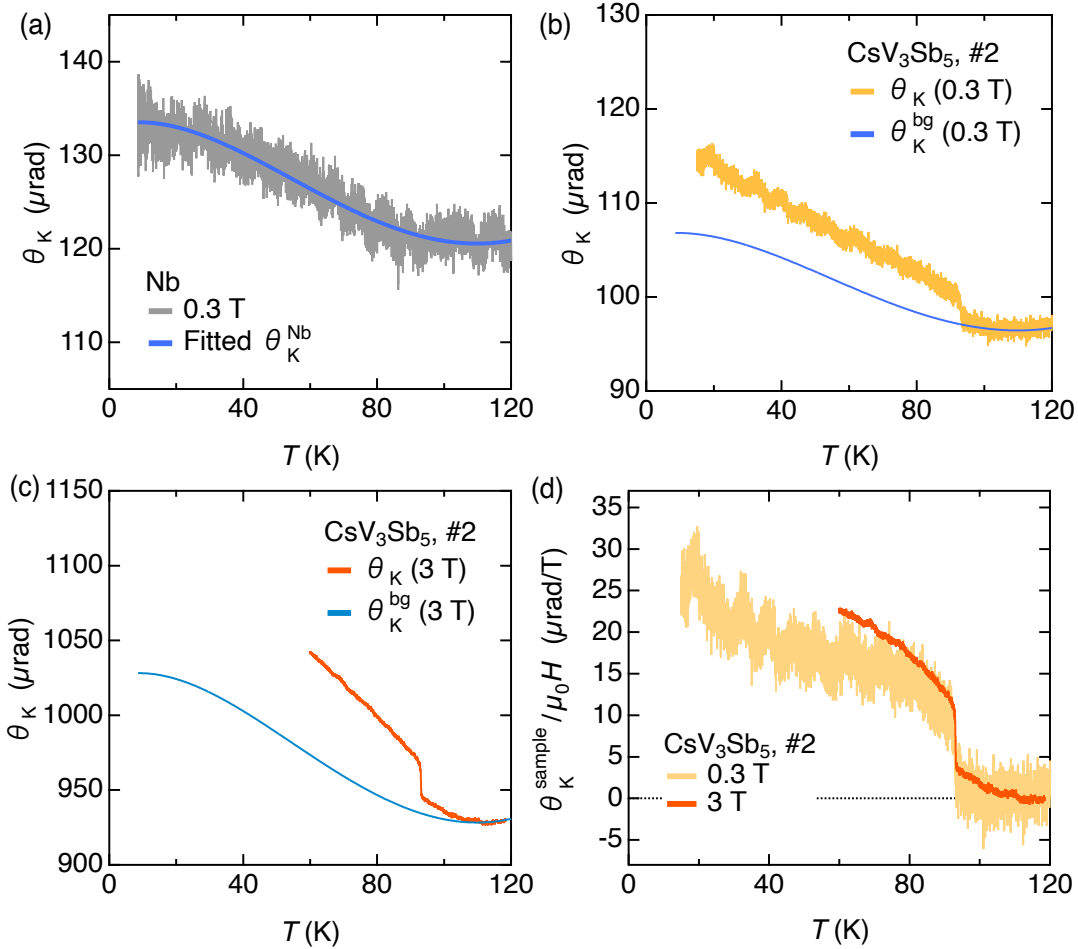

**Figure S2: Examples of background subtraction to produce the data in Fig 2a of the main text, for sample #2. a.** Temperature dependence of the polar Kerr angle $\theta_K$ of a Nb sheet measured at 0.3 T. The data is fitted by a polynomial function, as shown by the blue curve, $\theta_K^{Nb}$. **b.** Temperature dependence of $\theta_K$ of $CsV_3Sb_5$ measured at 0.3 T. The background signal is $\theta_K^{bg}$ (0.3 T) $= \theta_K^{Nb} \times 0.8$. **c.** Temperature dependence of $\theta_K$ of $CsV_3Sb_5$ measured at 3 T. The background signal is $\theta_K^{bg}$ (3 T) $= \theta_K^{Nb} \times 10 \times 0.77$. **d.** Temperature dependence of the polar Kerr angle after background subtraction $(\theta_K - \theta_K^{bg})$ divided by magnetic field, the same as data shown in Fig. 2a of the main text.

shows a drop, which is symmetric to the data shown in Fig. 2b measured under positive field.

In the temperature dependence of $\theta_K$, there is a tiny upturn (downturn) under positive (negative) magnetic field at $\sim$ 95 K (Fig. 2b of the main text), implying an onset behaviour. We calculated the temperature derivative of the polar Kerr angle $|d\theta_K/dT|$, as shown in Fig. S3c, in order to investigate this behaviour more in detail. The sharp peaks indicate the jumps in $\theta_K$ at the CDW transition. Above 5 T, $|d\theta_K/dT|$ shows a broad hump above $T_{CDW}$. This feature may be attributed to multiple domains within the light spot area, which is not surprising for a CDW system exhibiting a first-order phase transition. Moreover, the hump becomes larger at higher magnetic field. If the hump above $T_{CDW}$ comes from another domain which contribute only a tiny portion to the total signal, when the magnetic field is higher, $\theta_K$ is enhanced almost linearly and the hump part becomes more visible.

Figure S4 shows the data of sample #10. The temperature derivative of the second-harmonic signal $|dV_{2\omega}/dT|$ shows a peak around 93 K, as shown in Fig. S4a. The data are used to obtain $T_{CDW}$ up to 10 T (Fig. 1b). Figure S4b shows the temperature dependence of $\theta_K$. For opposite magnetic fields, both the $\theta_K$ and the jump at $T_{CDW}$ are flipped. The tiny kinks around 90 K in the $\pm 5$ T data are due to drastic temperature change. Figure S5 shows the similar data measured on sample #2. The $|dV_{2\omega}/dT|$ shows peaks at $T_{CDW}$ and the $\theta_K$ can be flipped by opposite magnetic fields.

In Fig. S6, we present the temperature dependence of $V_{2\omega}$ (proportional to reflectivity) near $T_{CDW}$ measured at 4 T with different light power. The data are normalized by $V_{2\omega}$ at 95 K. With

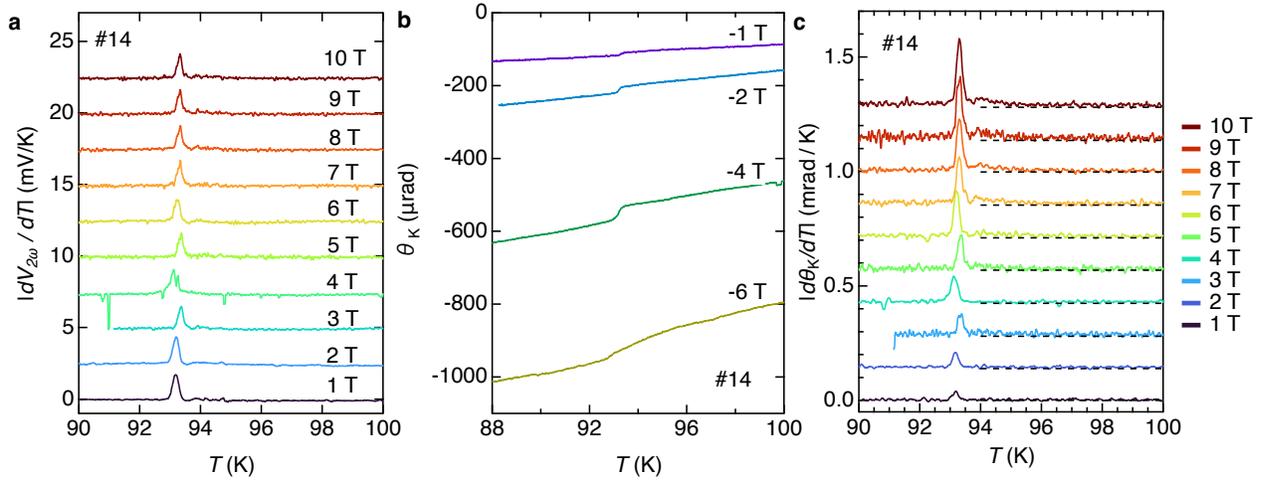

**Figure S3: Temperature derivative of $V_{2\omega}$, $\theta_K$, and its temperature derivative of sample #14.** **a**. Temperature dependence of the temperature derivative of $V_{2\omega}$ under magnetic field. The data are vertically offset for clarity. **b**. Temperature dependence of $\theta_K$ measured under negative magnetic field. **c**. Temperature dependence of $|d\theta_K/dT|$ under magnetic field. The data are offset for clarity. The dashed lines indicate the constant $|d\theta_K/dT|$ above $T_{CDW}$.

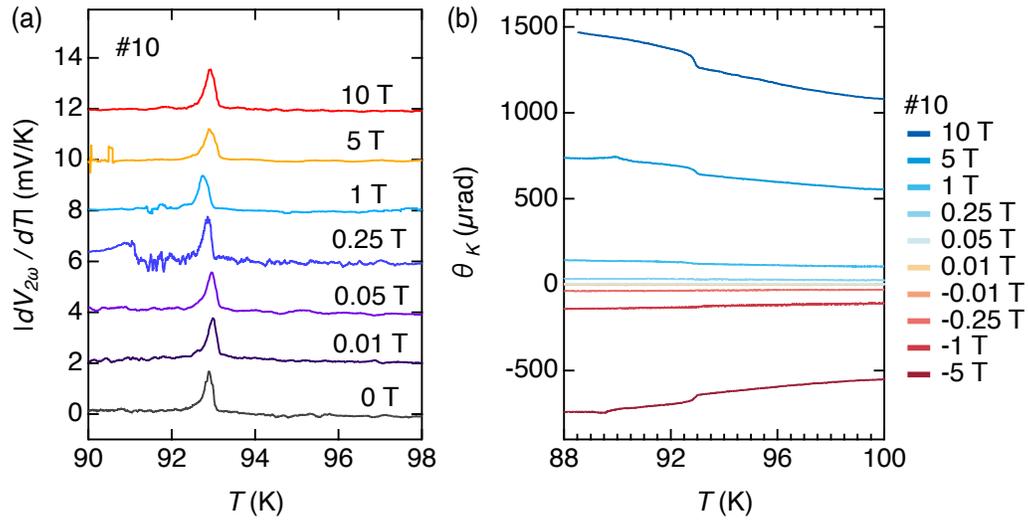

**Figure S4: Temperature derivative of $V_{2\omega}$ and $\theta_{\mathrm{K}}$ of sample #10.** **a.** Temperature dependence of the temperature derivative of $V_{2\omega}$ under magnetic field. The data are vertically offset for clarity. **b.** Temperature dependence of $\theta_{\mathrm{K}}$ measured under magnetic field.

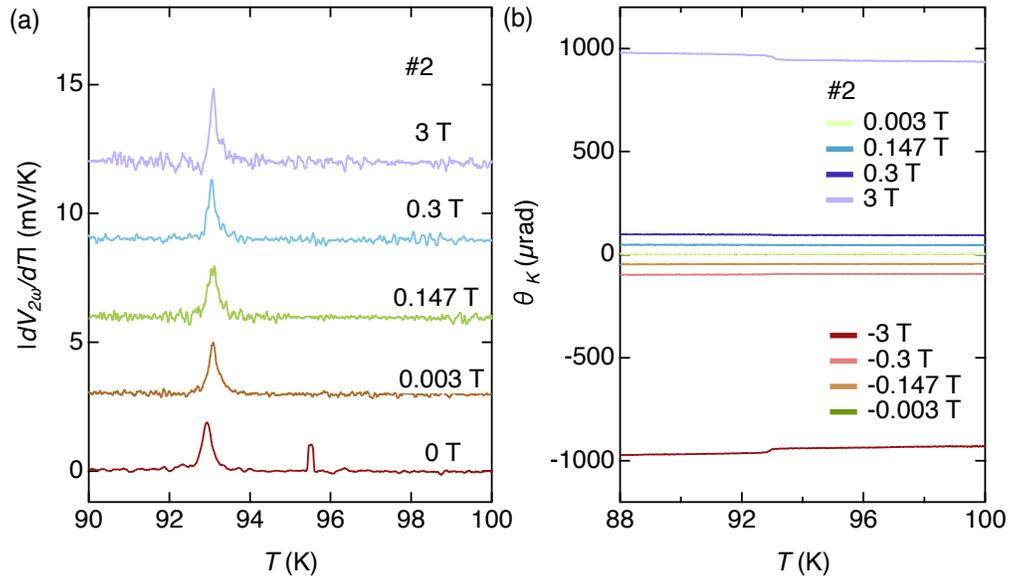

**Figure S5: Temperature derivative of $V_{2\omega}$ and $\theta_K$ of sample #2. a**. Temperature dependence of the temperature derivative of $V_{2\omega}$ under magnetic field. The data are vertically offset for clarity. **b**. Temperature dependence of $\theta_K$ measured under magnetic field.

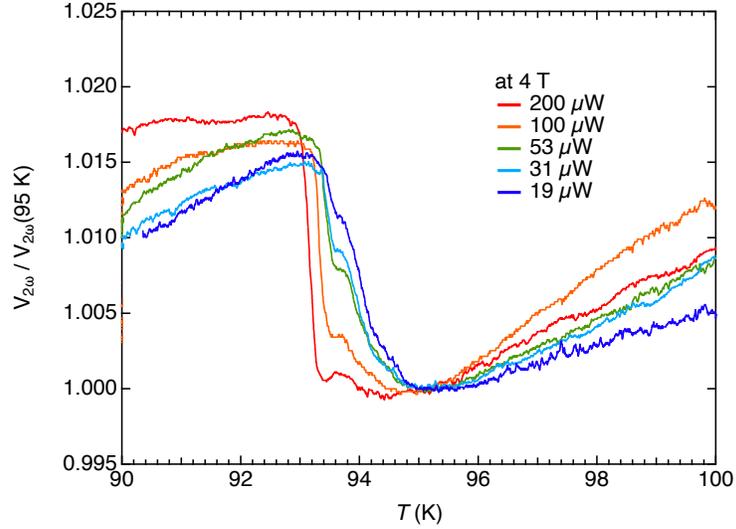

**Figure S6: Temperature dependence of $V_{2\omega}$ near $T_{\text{CDW}}$ with different light power.** The data are collected on sample #14 at 4 T. $V_{2\omega}$ is normalized to $V_{2\omega}(95 \text{ K})$ in order to compare the transition temperature.

decreasing power from 200 μW to 19 μW, the CDW transition temperature $T_{\text{CDW}}$ increases by around 1 K. Moreover, with smaller power, the transition tends to show two steps with a sharp jump at lower temperature and a broader jump at higher temperature. This feature is consistent with the domain effect and the change of effective size of the light spot. For a smaller power (smaller effective spot size), a fewer number of domains are measured and the contribution from minor domains (with higher $T_{\text{CDW}}$) becomes more visible. Comparing the sharp jump at the transition, the difference of $T_{\text{CDW}}$ between 200 μW and 100 μW is ∼ 0.2 K. Thus, the heating effect should not be severe in our measurements.



\end{document}